\documentclass[10pt,a4paper,twoside]{article}

\usepackage{indentfirst}			
\usepackage{graphicx}				
\usepackage{amsgen,amsfonts,amssymb,amsbsy}	

\newenvironment{resum}{\begin{quote}\small}{\end{quote}}

\def\r{\mathbb R}                   
\def\NAT{\mathbb N}                   
\def\z{\mathbb Z}                   
                   
\def\d{\partial}         
\def\fr{\frac}
\def\be{\begin{equation}}
\def\ee{\end{equation}}
\def\bea{\begin{eqnarray}}
\def\eea{\end{eqnarray}}
\def\b*{\begin{eqnarray*}}
\def\e*{\end{eqnarray*}}
\def\N{\hfill \raisebox{1mm}{\framebox{\rule{0mm}{1mm}}}}
\def\T{{\bf T}}
\def\S{{\bf S}}
\def\Q{{\bf Q}}
\def\G{{\bf g}}

\def\DP{{\cal DP}}

\def\C{{\cal C}}
\def\b{\bar}
\def\g{\gamma}

\def\Z{\Theta}
\def\z{\theta}
\def\O{{\bf\Omega}}
\def\f{\varphi}
\def\P{{\it Proof :} \hspace{3mm}}
\def\k{\vec{k}}

\def\xiv{\vec \xi}
\def\lie{{\pounds}_{\xiv}\, }

\def\u{\vec{u}}
\def\K{{\bf k}}
\def\xif{\mbox{\boldmath $ \xi $}}
\def\Diff{\mbox{\it Diff}}
\def\Conf{\mbox{\it Conf}}
\newtheorem{defi}{Definition}[section]
\newtheorem{theo}{Theorem}[section]

\newtheorem{prop}{Proposition}[section]

\newcommand{\bfsf}[1]{\textsf{\textbf{#1}}}

\begin{document}

\thispagestyle{plain}		

\begin{center}


{\LARGE\bfsf{Causal symmetries}}

\medskip


\textbf{A.\ Garc\'{\i}a-Parrado} and \textbf{J.M.M\ Senovilla}


\textsl{Universidad del Pa\'{\i}s Vasco, campus de Vizcaya (Spain)}

\end{center}


\begin{resum}

We define a new type of transformation for Lorentzian manifolds 
characterized by mapping every causal future-directed vector onto a causal
future-directed vector. The set of all such transformations, which we call 
causal symmetries, has the structure of a submonoid.
Some of their properties are investigated and we give
necessary conditions for a vector field $\xiv$ to be the infinitesimal generator 
of a one-parameter group of causal symmetries. Some examples are discussed.

\end{resum}


\section{Causal Tensors and Causal Relations}
Our basic arena is a smooth n-dimensional Lorentzian manifold $(V,\G)$ where $\G$ is
the metric tensor with signature $+,-,\ldots,-$. 
As usual, we denote by $T_{p}(V)$, $T(V)$ and ${\cal T}^{r}_s(V)$
the tangent space at $p\in V$, the
tangent bundle and the $r$-contravariant $s$-covariant
tensor bundle, respectively.  All our definitions can be
appropriately given for elements of ${\cal T}^{r}_s(V)$ or for 
their smooth sections. The hyperbolic signature
of $\G$ classifies each $\u\in T_p(V)$ as timelike, spacelike or null 
according to the sign of $\G (\u ,\u)$. The non-spacelike vectors are 
called {\em causal}. We assume that $(V,\G)$ is causally orientable
so that we can choose a smooth timelike vector field $\u_1$ as future pointing. 
Then, all causal vectors $\u_2$ split into future-directed and past-directed
according to whether $\G(\u_1,\u_2)> 0$ or $\G(\u_1,\u_2)<0$.
The symbol $\Z^+(V)$ will be used for the set of all
causal future-directed vectors of $V$. All the definitions and 
results have a dual replacing ``future'' by ``past'' and ``+'' 
by ``$-$'', which we will generally omit.
 
The splitting of vectors into causal and non-causal ones carries
over to higher-rank tensors as follows: $\T\in{\cal T}_r(V)$ is a 
{\em causal future-directed tensor} if $\T(\u_1,\dots,\u_r)\geq 0$ for 
all $\u_1,\dots,\u_r\in \Z^+(V)$. The set of $r$-covariant causal tensors 
is denoted by $\DP^+_r(V)$. This definition has obvious extensions
to tensor bundles other than ${\cal T}_r$.  The rank-2 case comprises
the tensors fulfilling the physically meaningful dominant energy condition
of General Relativity. The study of causal
tensors with $r\geq 2$ arose from the so-called ``super-energy'' 
tensors \cite{S-E1,S,S-E2}, and their interest has been analyzed 
in a number of works \cite{S-E3,S-E4}. Among their most interesting properties
we can mention their use in the proof of the causal propagation of
fields in very general settings \cite{BS} and the construction of conserved
currents \cite{S-E1,S}. For a given element $\T\in\DP^+_r(V)$ its
canonical directions are the elements $\k\in\Z^+(V)$ such that
$\T(\k,\dots,\k)=0$. These vectors are necessarily null and they are 
collected in a set denoted by $\mu(\T)$.

Here we are interested in the
role played by causal tensors in the study of causal mappings between
Lorentzian manifolds as they were introduced in \cite{CAUSAL}.
\begin{defi}
Let $(V,\G)$ and $(W,{\tilde \G})$ be Lorentzian manifolds and
$\f: V \rightarrow W$ a diffeomorphism. $\f$ is said to be a
{\em causal relation} if $\f^{'}(\Z^+(V))\subseteq\Z^+(W)$.
We will write this as $V\prec_{\f} W$. The Lorentzian manifolds
$(V,\G)$ and $(W,{\tilde \G})$ are {\em causally related} if such a $\f$ 
exists. This is denoted simply by $V\prec W$.
\end{defi}          
Next, we gather some important results proven in \cite{CAUSAL}
regarding causal relations:
\begin{theo}
$V\prec_{\f}W$ $\Longleftrightarrow$ $\f^{*}\tilde{\G}\in\DP^+_2(V)$ 
$\Longleftrightarrow$ $\f^{*}(\DP^+_r(W))\subseteq\DP^+_r(V)$ for all $r\in\NAT$
$\Longleftrightarrow$ $\f^{*}(\DP^+_r(W))\subseteq\DP^+_r(V)$ for a given odd
$r\in\NAT$.
\label{1ST}
\end{theo}
\begin{theo}
For a diffeomorphism $\f:(V,\G)\rightarrow (W,{\tilde\G})$ we have:\\ 
$V\prec_{\f}W$ with $\mu(\f^{*}\tilde{\G})=\d\Z^+(V)=$ future-directed null cone
$\Longleftrightarrow$ $\f^{*}{\tilde\G}=\alpha\G$, $\alpha>0$
$\Longleftrightarrow$ $(\f^{-1})^{*}\G=\beta{\tilde\G}$, $\beta>0$
$\Longleftrightarrow$ $V\prec_{\f}W$ and $W\prec_{\f^{-1}}V$.
\end{theo}
This theorems allows to put in causal correspondence two
different Lorentzian manifolds in a new fashion, which generalizes the 
conformal mapping, by means of reciprocal causal relations $V\prec_{\f}W$
and $W\prec_{\psi}V$ with $\psi\neq\f^{-1}$, see \cite{CAUSAL} for 
details.
\
\vspace{-5mm}
\ 
\section{Definition and properties of causal symmetries}
 From now on we work with one Lorentzian manifold $(V,\G)$. 
A smooth diffeomorphism $\f :V\rightarrow V$ is called
a {\em causal symmetry} if $V\prec_{\f}V$.  Denoting by $\Diff(V)$ 
and $\C(V)$ the set of transformations of $V$ and the set of causal 
symmetries, respectively, we derive from the previous theorems
that: (a) $\f_1\circ\f_2\in\C(V)$ for all $\f_1,\f_2\in\C(V)$ from where 
$\C(V)$ is a submonoid of the group $\Diff(V)$; and (b) the subgroup of 
units $\C(V)\cap\C(V)^{-1}$ is precisely the set of all conformal 
transformations $\Conf(V)$.

We now try to define continuous substructures of causal symmetries in much the
same way as with the continuous subgroups of isometries or
conformal transformations.
Note, however, that the inverse of a causal symmetry is not a causal symmetry
unless it is a conformal one.  Thus if $I\subset\r$ is a connected
interval containing $0$ and $\{\f_s\}_{s\in I}$ is a local
one-parameter group of diffeomorphisms we get \cite{SYMMETRY}:
\begin{theo}\begin{enumerate}
\vspace{-1mm}
\item $[0,\epsilon)\subseteq I$ and
$\{\f_s\}_{s\in[0,\epsilon)}\subset\C(V)\Longrightarrow\{\f_s\}_{s\in\r^+\cap I}
\subset\C(V)$.
\vspace{-3mm}
\item $\{\f_s\}_{s\in I\cap\r^+}\subset\C(V)$ and 
$\f^{'}_{s_0}\k\in\d\Z^+$ ($s_{0}>0$)
for $\k\in\d\Z^+\Longrightarrow\f^{'}_s\k\in\d\Z^+$.
\item $\f_{s_0}\in \Conf(V)$ ($s_{0}>0$) and
$\{\f_s\}_{s\in[0,\epsilon)}\subset\C(V) \Longleftrightarrow \{\f_s\}_{s\in I}
\subset \Conf(V)$.
\end{enumerate}
\end{theo}  
This theorem implies that we can only speak of
one-parameter {\em submonoids} of causal symmetries---unless {\em all} 
of them are conformal transformations, in which case it is a 
conformal subgroup---, and they are isomorphic to half the real line
in the global case (and to $\r^+\cap I$ in the local one). 

\
\vspace{-5mm}
\ 
\subsection{Canonical null directions}
The canonical null directions of a submonoid of causal symmetries are the null
vector fields which remain null under the push-forward of
$\{\f_s\}_{s\in I\cap\r^+}$. We collect them in the set $\mu(\f_s)$, 
and one can easily show $\mu(\f_s)=\mu(\f^*_s\G)$.
The important point is that $\mu(\f_s)$ does not actually depend on $s$
and is intrinsic to the one-parameter submonoid \cite{SYMMETRY}.
Of course $\mu(\f_s)$, being a set of null directions, is not a vector space.
Nevertheless, we can pick up a maximal number of {\em linearly 
independent} null vector fields, say $\k_1,...,\k_r$, of
$\mu(\f_s)$ and the vector space $Span\{\k_1,...,\k_r\}$ has the 
property of remaining invariant under the push-forward of the submonoid.
The number $r$ is also intrinsic to $\mu(\f_s)$. Let $\K_1,...,\K_r$ be
the 1-forms associated ---``lowering indices''--- to the vectors
$\k_1,...,\k_r$ and define $\O=\K_1\wedge\dots\wedge\K_r$. Then we have
\cite{SYMMETRY}
\begin{theo}
$\f^{*}_s\O=\sigma_s\O$ for some $\sigma_s\in C^{\infty}(V)$, $s\in I$. 
\label{DIRECTIONS}
\end{theo} 
In the cases $r=1,2$, $\f^{'}_s\k\propto\k$ for every $\k\in\mu(\f_s)$.
A straightforward consequence of theorem \ref{DIRECTIONS} is that
$\lie\O\propto\O$ where $\xiv$ is the generator of the local
one-parameter group $\{\f_s\}_{s\in I}$.  Another important result is the next
proposition \cite{SYMMETRY}:
\begin{prop}
 For every $\T\in\DP^+_r(V)$ and $\{\f_s\}_{s\in\r^+\cap I}\subset\C(V)$  
$$
\f^{'}_s(\mu(\f^*_s\T))=\mu(\T)\cap\mu(\f_s),\ s\in\r^+\cap I .
$$
\end{prop} 
Hence, for instance, if $\T$ does not share canonical null directions with
$\{\f_s\}$ then $\mu(\f^*_s\T)$ is empty for $s>0$, while if $\mu(\f_s)$ is
a subset of $\mu(\T)$ then $\mu(\f^*_s\T)=\mu(\f_s)$.
\
\vspace{-5mm}
\ 
\subsection{Necessary conditions for infinitesimal causal symmetries}
Let us now present the necessary conditions for the infinitesimal generator
$\xiv$ of a submonoid of causal symmetries. The main result gives the 
deformation of the metric tensor under the action of $\xiv$:
\begin{theo}
If $\{\f_s\}_{s\in\r^+\cap I}\subset\C(V)$ then
$\lie\G-\alpha\G\in\DP^+_2(V)$ for some smooth function $\alpha$ in
$V$.  Furthermore $\mu(\f_s)=\mu(\lie\G-\alpha\G)$.
\label{DECOM}
\end{theo}
\P According to theorem \ref{1ST}, $\f^{*}_s\G\in\DP^+_2(V)$ for
$s\in\r^+\cap I$ so that we can apply the canonical
decomposition for elements of $\DP^+_2(V)$ proven in \cite{S-E2}.
Roughly speaking, this decomposition splits every rank-2
causal tensor in a sum of super-energy tensors of 
$p$-forms $\O_{[p]}$ (this is written as $\T\{\O_{[p]}\}$)\footnote{We can think
of each of these super-energy
tensors as the electromagnetic energy-momentum tensor constructed out
from $\O_{[p]}$ as the Faraday tensor.},
where the $\O_{[p]}$ are simple and causal, i.e. they are the wedge
product of null 1-forms. Thus,    
\[
\f_s^{*}\G=\sum_{p=r}^{n-1}\T\{\O_{[p]}(s)\}+A_s^2\G 
\]
where we have used that $\T\{\O_{[n]}\}\propto\G$. 
As $A_{0}=1$ it follows, due to the continuity in $s$, that $A_{s}\neq 0$ for
$s\in (0,s_{0})$, so that for small enough $s$ we can derive
\[  
f_{\u_1,\u_2}(s)\equiv A^{-2}_{s}\f^*_{s}\G(\u_1,\u_2)\geq \G(\u_1,\u_2)
\]
for all $\u_1,\u_2\in\DP^+_1(V)$.
Therefore $f^{'}_{\u_1,\u_2}(0)=(\lie\G-\alpha\G)(\u_1,\u_2)\geq 0$
$\forall \u_1,\u_2\in\DP^+_1(V)$ proving the first
statement of the theorem.  For the second one, by letting
$(\f^{*}_s\G)(\k,\k)=0$ we get at once $(\lie\G-\alpha\G)(\k,\k)=0$
and hence $\mu(\f_s)\subseteq\mu(\lie\G-\alpha\G)$.  The inclusion
$\mu(\lie\G-\alpha\G)\subseteq\mu(\f_s)$ is a little bit more involved,
see \cite{SYMMETRY}.\N

 Using this result we can apply the decomposition theorem once more
to get $\lie\G-\alpha\G=\beta\, \S+\Q$
where $\beta>0$, $\Q\in\DP^+_2(V)$ is a remainder with
$\mu(\f_s)\subseteq\mu(\Q)$, and $\S\propto\T\{\O\}$ is the normalized 
superenergy tensor of the simple $r$-form $\O$ appearing in theorem
\ref{DIRECTIONS}. This normalization is achieved by choosing $\O$ such that 
$\O\cdot\O=2r!(-1)^{r-1}$ for $r>1$ (the case $r=1$ is degenerate and 
has $\O=\K$), so that $S_{ap}S^{p}_{\ b}=g_{ab}$ for $r>1$ ---or
$S_{ap}S^{p}_{\ b}=0 (\Longleftrightarrow \S=\K\otimes\K)$
for the degenerate case--.
Since $\lie\O$ is known (theorem \ref{DIRECTIONS}) we can work out the value
of $\lie\S$, thus obtaining: 
\begin{eqnarray*}
\lie\O=\fr{1}{2}r(\alpha+\beta+\lambda)\O &\Longleftrightarrow& \lie S_{ab}=
\beta g_{ab}+\alpha S_{ab}+Q_{ap}S^{p}_{\ b} \,\,\, (r>1) , \\
\lie \K=\g \K &\Longleftrightarrow& \lie S_{ab}=2\g S_{ab}\,\,\, (r=1).
\end{eqnarray*}
These equations together with that of $\lie\G$ are
necessary conditions for the vector $\xiv$ to generate a submonoid of
causal symmetries. We believe they are also sufficient, and we have 
been able to prove this explicitly in the following situations:

\noindent
{\bf Case 1} When $r=n-1$. Then there are $n-1$ independent canonical null
directions and the remainder $\Q$ is just a term proportional to the
metric tensor. Consistency conditions between the Lie derivative and
the algebraic properties of $\S$ and $\G$ lead to
$\lie\G=\alpha\G+\beta\S ,\ \ \ \lie\S=\beta\G+\alpha\S$.

\noindent
{\bf Case 2} For the {\em pure} cases when the remainder $\Q$ vanishes.
Then we get the same equations as in case 1 for $r>1$,
and $\lie\G =\alpha\G+\beta\K\otimes\K$, $\lie \K=\gamma \K$ for
the degenerate case $r=1$. This last possibility includes the 
so-called Kerr-Schild vector fields and symmetries recently considered 
in \cite{KERR-SCHILD}. 

In both cases, the integration for $\f^{*}_s\G$ provide solutions 
which are easily shown to be causal tensors if $s>0$ thus
generating a submonoid of causal symmetries. The
parameters $\alpha,\beta$ and $\g$ appearing in the equations
are known as the gauges of the symmetry \cite{KERR-SCHILD},
and of course they depend on $\xiv$. Moreover, if $\xiv$ and $\xiv_1$ 
are two generators of submonoids of causal symmetries, then the Lie commutator
$[\xiv,\xiv_1]$ will also comply with the necessary conditions but now
$\beta_{[\xiv,\xiv_1]}$ need not be positive and hence $[\xiv,\xiv_1]$
will not in general generate another submonoid of causal symmetries.  
Thus, if one wants to keep the Lie algebra structure, the set of vector 
fields must be enlarged to include gauge functions $\beta$ of any sign.
\ 
\vspace{-5mm}
\ 
\section{Examples}
\noindent
{\bf Example 1} Let us consider Vaidya spacetime given by the line element:
$$
(V,\G):ds^2=\left(1-2M(t)/r\right)dt^2-2dtdr-r^2(d\z^2+\sin^2\z d\f^2),
$$
and define the one-parameter group acting over $V$ as $\f_s:t\mapsto
t+s$.  Then
$$
\f^{*}_s\G=\G +[M(t)-M(t+s)]/r \,\, dt\otimes dt,
$$
from what we conclude that this tensor is a causal one if $M(t)$ is a
decreasing function.  The infinitesimal generator $\xiv=\d/\d t$
satisfies
$$
\lie dt=0,\ \ \lie\G=-\dot{M}(t)/r\,\, dt\otimes dt,
$$
which corresponds to the degenerate case of a single canonical null direction
given by $\mu(\f_s)=\{\d/\d r\}$, see also \cite{KERR-SCHILD}.
This vector field is tangent to the outward incoherent radiation and 
defines a time arrow which is preserved by the submonoid.

\noindent
{\bf Example 2\ }  In this example we give a very simple causal symmetry with
$n-1$ canonical null directions. The line-element is given by
$$
(V,\G):ds^2=F^2(x^{n-1})\sum_{i,j=0}^{n-2}g_{ij}(x^i)dx^{i}dx^{j}-(dx^{n-1})^2,
$$
and the causal mappings by $\f_s:x^{n-1}\mapsto x^{n-1}+s$.
It is straightforward to verify that $\{\f_s\}_{s\in\r^+}$
is a submonoid of causal symmetries if $F^2(x^{n-1}+s)\geq F^2(x^{n-1})$.
The infinitesimal generator $\xiv =\d/\d x^{n-1}$ obeys the relations
$$
\lie \G=2F^{'}(x^{n-1})(\G+\xif\otimes\xif ),\,\,\, \lie\xif =0,
$$
which are equivalent to those of case 1 for
$\O\propto dx^0\wedge\dots\wedge dx^{n-2}$.


\end{document}